\documentclass[]{aa} 

\usepackage{graphicx,amsmath,amssymb}

\begin{document}

\thesaurus{(13.07.1 ; 09.03.2 ; 02.01.1 ; 02.09.1)}

\title{Acceleration of UHE Cosmic Rays in Gamma-Ray Bursts} \author{Guy
  Pelletier\inst{1,2} \and Evy Kersal\'e\inst{1} }

\institute{Laboratoire d'Astrophysique de l'Observatoire de Grenoble,
  BP 53, F-38041 Grenoble cedex 9, France \and Institut Universitaire
  de France }

\offprints{Guy.Pelletier@obs.ujf-grenoble.fr}

\date{Received April 2000 / Accepted May 2000}

\maketitle

\begin{abstract}
  Gamma-Ray Bursts are good candidates of the ``bottom up'' scenario
  for the generation of the UHE Cosmic Rays. In the most discussed
  model of GRBs, namely the ``fireball'' model, a highly relativistic
  shock forms and seems capable of accelerating the cosmic rays up to
  the EeV range. However, only the first Fermi cycle produces a large
  energy gain to particles coming from the external medium. Thus, a
  complementary acceleration is proposed, downstream of the external
  shock, in the relativistic plasma of the GRBs, where crossings of
  relativistic fronts are likely to occur.  Both forward and backward fronts
  are necessary for the internal Fermi acceleration to work and the
  physical process that generates them is presented. We found that
  there exists a relevant physical process similar to Brillouin
  backscattering that redistributes the incoming energy in the plasma
  shell. This redistribution occurs through the generation of sound
  waves that heat the plasma shell and also through the generation of
  both forward and backward relativistic Alfv\'en fronts that accelerate
  cosmic rays by the Fermi process. We show that this ensemble of
  processes is able to account for the generation of UHE cosmic rays.
    
  There are two opportunities for these combined processes, first
  during the ``primary'' Gamma-Ray Burst where baryons are entrained
  by the relativistic pair wind, second during the predeceleration and
  deceleration stages when the fireball interacts with the
  interstellar medium. However, because of the synchrotron losses,
  only the second stage can produce the UHE cosmic rays.
  

\end{abstract}

\section{Introduction}\label{sec.1}
The origin of the cosmic ray population beyond $10^{15}eV$ is an
important enigma of modern astrophysics which will hopefully be
clarified by the Pierre Auger instrument. Beyond this energy, the
galactic magnetic field is unable to scatter cosmic rays and moreover
supernovae remnants cannot accelerate protons.
Pulsars could accelerate protons up to $10^{17}eV$. This population,
that has an isotropic energy spectrum with a powerlaw index $3.1$,
probably comes from extragalactic sources. The ``knee'' of the
spectrum around $10^{15}eV$ is so smooth that the extragalactic
contribution could be important at lower energy also.

The photoproduction of pions by cosmic ray protons on the Cosmic
Microwave Background, the GZK-effect, occurs beyond the energy
threshold of $3 \times 10^{19}eV$ and higher energy protons cannot
come from sources located farther than $100 Mpc$ (Aharonian \& Cronin
\cite{aha}). Few events beyond the GZK-threshold have recently been
detected by AGASA (Hayashida et al. \cite{aga}) and the ``Fly's eyes''
(Sokolsky \cite{fly}), and the Auger experiment will considerably
improve the statistics of these events. These preliminary results
suggest that a new population of cosmic rays is observed at energies
higher than $10^{18}eV$, because the spectrum becomes harder.  The
origin of these few events, although some pointing ability of the
instruments, remains very uncertain. The identified extragalactic
sources within $100 Mpc$ are very few.

Vietri (\cite{vie}) and Waxman (\cite{wax}) showed that the rate of
GRBs in the Universe ($10^{-8} yr^{-1} Mpc^{-3}$) and their energy
($10^{51}-10^{53} erg$) make them very good candidates as sources of
the UHECRs, if $10 \%$ of their energy is converted into UHECR energy.
The most considered GRB model, namely the ``fireball'' model (Rees \&
M\'esz\'aros \cite{ree}), is based on a relativistic blast wave having a
Lorentz factor $\Gamma_{s}$ larger than $10^{2}$, this high value
being necessary for gamma photons to escape from pair production. A
Fermi cycle at a relativistic shock can amplify the energy of cosmic
ray by a factor $\Gamma_{s}^{2}$. Thus two Fermi cycles could be
sufficient to reach the expected energy of the UHECRs. However,
Gallant \& Achterberg (\cite{gal}) recently showed that only the first
cycle can produce such amplification, the next cycles amplifying by a
factor of 2 only. Moreover the escape probability is large ($0.3-0.5$).

The GRB light curves indicate strong internal disturbances and they
have been considered as the main cause of gamma-ray emission and a
possible cause of cosmic ray generation (Waxman \& Bahcall \cite{wax2}
; Daigne \& Mochkovitch \cite{daim}).  In this paper we propose an
interpretation of these disturbances in terms of relativistic
hydromagnetic fronts, calculate their generation through the streaming
instability caused by baryon loading, show that there is an efficient
nonlinear generation of both forward and backward waves, estimate the
amount of energy they extract from the fireball and calculate the
acceleration of cosmic rays by their crossings. The baryon loading
occurs first during the ``primary'' GRB when the relativistic pair
wind entrains the debris of the progenitor, and then during the
interaction of the fireball with the interstellar medium. Furthermore,
the first loading is likely to occur in the ``hypernova'' scenario,
however, this is less obvious in the ``merging'' scenario.

\section{Loading baryons}\label{sec.lb}

The canonical fireball model (Rees \& M\'esz\'aros \cite{ree}) presents
a primary explosion during which the plasma is optically thick until
$t_*$ ($R_* = ct_*$), then an adiabatic and free expansion until the
ambient medium starts to exert sufficient ram pressure; this defines
a deceleration time $t_d$ and a deceleration radius $R_d$ (typically
$t_* = 3 \times 10^{-3} t_d$). The third stage is thus the
deceleration one during which most of the energy of the fireball is
dissipated in the interstellar medium, giving rise to the afterglow.
We assume that the fireball is initially composed of $e^{+}-e^{-}$
pairs and that this ultra-relativistic wind of bulk Lorentz factor
$\Gamma$, flowing along a strong poloidal magnetic field, is perturbed
by ambient baryons of mass density $\rho_{b}$.  There are two stages
of baryon loading, during the ``primary'' GRB and during the
interaction with the interstellar medium.

\subsection{Hydrodynamics of the baryon loading stage}\label{subsec.hyd}

The main features of the hydrodynamics of the relativistic fireball can be
derived from the energy invariant as long as the radiation losses are
negligible (assuming radial magnetic field lines):
\begin{equation}
        E = \int (e+\beta^{2}P) \gamma^{2} d^{3}v \ ,
        \label{ENI}
\end{equation}
where $e$ is the comoving energy-mass density, $P$ the comoving
pressure, $\beta$ the flow motion (in unit of light velocity), and
$\gamma$ the Lorentz factor of the flow ($\gamma \equiv
(1-\beta^2)^{-1/2}$), distributed around the typical bulk Lorentz
factor $\Gamma$.  We assume that the plasma shell is composed of a low
pressure baryonic component of mass $M$ and a high pressure
relativistic component containing pairs and cosmic rays of average
energy density in comoving frame $e_{*}=3P$. We approximate the energy
by the following expression, assuming an appropriate self-similar
evolution of the shell (which allows us to properly define the bulk
Lorentz factor):
\begin{equation}
        E=M \Gamma c^{2} + 4 P \,\Gamma V_{0} \ ,
        \label{ENIA}
\end{equation}
where $V_{0} = 4\pi R^{2} \delta R_{0}$ is the covolume of the shell, $R(t)$
the shell radius, $\delta R_{0}$ its comoving thickness.  During the
expansion, the energy in the shell is redistributed between kinetic and
internal energies through the competition of adiabatic cooling and heating by
the flux of incoming matter. Part of the mass flux increases the low pressure
baryonic mass: $\dot M = (1-\alpha) \rho_{b} \Gamma c 4\pi R^{2}$; the other
part contributes to the high energy component: $\dot e_{*}\mid_{heating} =
\alpha \rho_{b} \Gamma c^{3} 4\pi R^{2}/V_{0}$, assuming a heating length
shorter than the fireball width. The adiabatic cooling is such that $\dot
e_{*}\mid_{adiab} = -\frac{4}{3}\frac{\dot V_{0}}{V_{0}} e_{*}$. Thus setting
$\dot E = 0$, after some simple manipulations, we obtain (with $\alpha_{0}
\equiv 1+\alpha/3$):
\begin{equation}
        \alpha_{0} \rho_{b} 4\pi R^{2}c^{3}\Gamma^{2} + \frac{1}{3}
        \frac{\dot V_{0}}{V_{0}}M\Gamma c^{2} + (\frac{\dot \Gamma}{\Gamma}
        - \frac{1}{3} \frac{\dot V_{0}}{V_{0}})E = 0 \ .
        \label{EVE}
\end{equation}
It can easily be checked that, when there is no mass input, an adiabatic
expansion occurs such that $\Gamma = \Gamma_{0}(V_{0}(t)/V_{0}(t_{0}))^{1/3}$
as long as $M\Gamma c^{2} \ll E$, followed by a free expansion $\Gamma \simeq
E/Mc^{2}$ when the internal energy has become smaller than the kinetic energy.
Now to analyze the effect of the mass sweeping, we assume a self-similar
expansion such that $d\ln V_{0}/d\ln t = \chi$ and define the dimensionless
quantity $\eta(t)$ (within a coefficient of order unity) as the ratio of the
fireball energy over the swept energy-mass of the ambient medium, measured in
the comoving frame, such that
\begin{equation}
        E = \eta(t) \rho_{b}\frac{4}{3}\pi R^{3}\Gamma^{2} c^{2} \ .
        \label{CHI}
\end{equation}
Then equation
(\ref{EVE}) leads to the differential equation:
\begin{equation}
        \frac{d \eta}{d\ln t} = 2 \eta (\eta_{\infty} - \eta) \ ,
        \label{ECHI}
\end{equation}
where $\eta_{\infty}$ is the asymptotic value taken by $\eta(t)$, namely
$\eta_{\infty} = \frac{3}{2}+\frac{\chi}{3}$.  With $\eta(t_{0}) = \eta_{0}$,
the solution is
\begin{equation}
  \eta(t) = \eta_{\infty}
  \frac{\eta_{0}(t/t_{0})^{2\eta_{\infty}}}{\eta_{\infty}-\eta_{0}
    +\eta_{0}(t/t_{0})^{2\eta_{\infty}}} \ .
  \label{CHIT}
\end{equation}
The evolution starts in the adiabatic regime, $\eta(t) \simeq
\eta_{0}(t/t_{0})^{2\eta_{\infty}}$ and $\Gamma \propto t^{\chi/3}$.
If $\eta$ reaches its asymptotic value, then the bulk Lorentz factor
decreases according to the canonic law $\Gamma \propto t^{-3/2}$.
However, the interest of this derivation is to stress that this
Blandford \& Mc Kee (\cite{blan}) law does not mean that a free
expansion has been reached and a strong relativistic shock set up.
Indeed the energy of the fireball can still be dominated by the
internal energy, so that a shock does not form, and the incoming
baryons interact with the fireball plasma through a streaming
instability. This is probably what happens during the primary stage of
baryon loading where a high entropy pair plasma entrains some baryonic
mass, expected to be of the order $10^{-6}M_{\odot}$, coming from the
debris of the progenitor. This interaction is probably more important
than the final stage of interaction with the interstellar medium in
generating perturbations in the flow as described in subsection
\ref{sec.relax} and section \ref{sec.bb}.

\begin{figure}[h]
  \includegraphics[width=0.8\columnwidth]{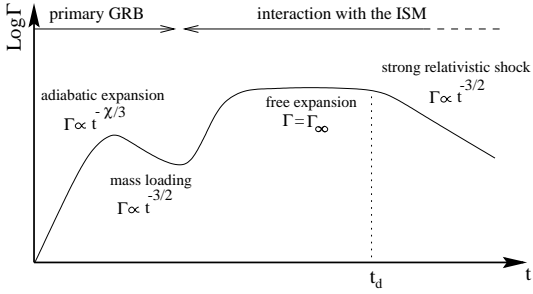}
    \caption{Evolution in time of the Lorentz factor $\Gamma$. The
      deceleration giving rise to the afterglow starts at time $t_d$,
      after a free expansion involving the asymptotic bulk Lorentz
      factor $\Gamma_{\infty} = E/Mc^2$.}
    \label{fig:gamma_t}
\end{figure}

Thus there are two stages of baryon loading (see
Fig.~\ref{fig:gamma_t}), the primary stage when the high entropy pair
wind entrains a mass $M_{0} \sim 10^{-6}M_{\odot}$ coming from the
debris of the massive star giving rise to the hypernova or of the
merging compact objects, and a secondary stage corresponding to the
sweeping of the ambient interstellar medium but that loads typically
$10^{-3}M_{0}$ only.

\subsection{The magnetic field and some scales}
\label{sec.mfs}

 We assume that this
baryonic plasma flows along a strong poloidal magnetic field. This is the
major assumption of this paper: $B>B_{eq}$ (we note $\beta_{p} \equiv
(B_{eq}/B)^{2} <1$) and $B>B_{m}$ where $B_{eq}$ is the equipartition field
corresponding to the plasma pressure in the fireball, and $B_{m}$ is the
equipartition field corresponding to the ram pressure exerted by the incoming
baryons.
\begin{equation}
    B_{m} \simeq 10^{2} (\frac{n_{b}}{1 \ cm^{-3}})^{1/2}
    \frac{\Gamma}{10^{3}} \ Gauss \ .
    \label{eq:BM}
\end{equation}
Indeed, the poloidal magnetic pressure follows the same decrease as
the relativistic pressure with distance; for conical expansion,
$B_{p}^{2}/8\pi \propto P \propto R^{-4}$. Moreover, a toroidal
magnetic field is likely significant in the edge of the collimated
flow and can play a confining role.  The decay of the toroidal field
is weaker, because $B_{t} \propto R^{-1}$ in a conical expansion.  If
the magnetic field dominates at the beginning, it remains so with the
expansion. Moreover, a tiny initial toroidal component of the magnetic
field can become dominant in the expansion and favors collimation.
Like in extragalactic jets, the poloidal field can be dominant along
the axis of the flow and the toroidal field dominant in the edge of
the flow. Thus we assume an internal poloidal field such that $B_p =
B_d(R/R_d)^{-2}$ with $B_d\gtrsim B_m$ at $R_d \sim 10^{16}cm$ and an
external toroidal field $B_t \gtrsim B_d(R/R_d)^{-1}$; in particular
at $R_*$ (typically $10^{-3}R_d$), $B \sim 10^8 G$. Of course these
estimates are sensitive to the model of magnetic field distribution,
however the numbers we proposed are reasonable and help to explain out
what happens.

In the early stage the proton energy is severely limited by
synchrotron losses. Assuming an acceleration time $t_{acc} = \kappa
\,t_L$, where $t_L$ is the Larmor time, the maximum proton energy is
such that the acceleration time is equal to the synchrotron time and
gives
\begin{equation}
  \epsilon_{max} = 2 \frac{10^{11}}{\sqrt{\kappa}} \left( \frac{B}{1 G}
  \right)^{-1/2}GeV.
  \label{eq:cutoff}
\end{equation}
Taking $\kappa \simeq 10$, the energy of $10^{17} eV$ can be achieved
only in the region where $B$ is smaller than $B_s \simeq 10^6 G$,
which requires $R>R_s \simeq 10^{-2}R_d$. Bearing in mind that
$\Gamma$ reaches a value between $10^2$--$10^3$, the UHECRs ($\sim
10^{20} eV$ in the observer frame) are necessarily produced in the
region $R>R_s$.

A neutrino radiation is produced during the early stage through
pp-collisions thanks to Fermi acceleration.  Indeed, $\epsilon_{max}$
given by (\ref{eq:cutoff}) exceeds $1GeV$ when $B<10^{12}G$, therefore
where $R>R_h \sim 10^{-2}R_*$. This low energy neutrino emission ends
when the pp mean free path becomes larger than $\delta R$, thus when
$R=R_{pp}\sim 10^2 R_h\sim R_*$. This stage of Fermi acceleration
supports the predictions made by Paczynski \& Xu (\cite{pac}), namely
a neutrino emission between $30 \,GeV$ -- $TeV$ of global energy of a
few percent of the fireball energy.

\subsection{Relaxation of the baryon stream in the fireball}
\label{sec.relax}

In this subsection, we examine how the backstream of baryons in the
fireball, initially composed of $e^{+}-e^{-}$ pairs, undergoes a
relaxation by triggering a beam instability in the pair plasma.

The hydromagnetic perturbations so produced scatter the baryons that
are then entrained and some of them accelerate to high energy. We
presume that these perturbations are those revealed by the light curve
and which also accelerate the electrons responsible for the
synchrotron and inverse Compton gamma-emission. Indeed, initial
perturbations do not amplify in the expansion, they even decay (this
differs from an expanding universe that is self-gravitating).
Therefore we stress that these perturbations must be generated during
the expansion by the appropriate instability.

At the presumed energy of the particles, the Coulomb interactions are
negligible (indeed, for relativistic electrons, the Coulomb cross
section is as low as the Thomson cross section; as long as the shell
is optically thin to Compton scattering, the mean free path is larger
than the shell width; this is of course more obvious when protons are
involved in the collision).  Therefore the incoming baryons interact
only with the magnetic field carried by the fireball.  The magnetic
field represents a more efficient obstacle if it is perpendicular to
the flow, and the interaction starts in the form of an intense
backward fast magneto-sonic wave. In a confined relativistic plasma,
the fast mode propagates with a Lorentz factor $\gamma_{F} =
\gamma_{S} \gamma_{*}$, where the sound Lorentz factor (corresponding
to the parallel slow mode actually) $\gamma_{S} = \sqrt{3/2}$ and the
Lorentz factor of the generalized Alfv\'en waves (see Pelletier \&
Marcowith \cite{pelmar}) $\gamma_{*} = (1+
\frac{1}{2\beta_{p}})^{1/2}$.  However, since we consider a collimated
expansion, the most natural circumstance is the interaction along the
parallel magnetic field within a wide solid angle around the axis.

In this scope, the interaction scenario is the following. In the
comoving frame, the proton back-stream, of velocity $v_{b}=\beta_{b}c$
and Lorentz factor $\Gamma_{b}$, along the field line generates
backward Alfv\'en waves of velocity $V_{*}=\beta_{*}c$ at a rate $g$
given by (Marcowith et al. \cite{mar} ; Pelletier \& Marcowith
\cite{pelmar}):
\begin{equation}
  g = g_{0}\,\frac{\Gamma}{\Gamma_{b}}\sqrt{\beta_{b}-\beta_{*}} \ ,
  \label{gro}
\end{equation}
where $g_{0} \equiv \frac{\omega_{pi}}{\Gamma}(2 \beta_{*})^{-1/2}$,
$\omega_{pi}$ being the plasma frequency of the ions in the stream.
This maximum growth occurs for the wave number
\begin{equation}
  k_{0} = \frac{\omega_{ci}}{\Gamma_{b} (v_{b}-V_{*})} \simeq
  \frac{\gamma_{*}^{2}}{\Gamma_{b}}\frac{\omega_{ci}}{c} \ .
  \label{kzo}
\end{equation}
 
When these backward waves have reached a high level, they scatter the
baryons that are then entrained by the pair flow. Thus, we presume that
the large internal disturbances observed in the light curve result
from loading baryons in the fireball either at the early stage of the
explosion or during the ``free'' expansion stage. At the beginning of
the fireball expansion in the interstellar medium, the diffusion
length is not short enough compared to the narrow width of the shell.
Then baryon entrainment starts when the level of the resonant Alfv\'en
waves (hereafter A-waves) is strong enough to get a diffusion length
shorter than the shell thickness. Then the kinetic flux of the
incoming baryons is transformed into a backward flux of intense
A-waves.  As long as these waves do not interact with the plasma shell
(their wavelengthes are larger than the dissipation scale), the shell
does not decelerate and just an electromagnetic wake is generated
propagating towards to shell center but more slowly than the
ultrarelativistic expansion of the shell so that it seems to advance.
Thus for a baryon mass loading rate $\dot M_{b} = 4\pi R^{2} \rho_{b}
\Gamma c$, the energy flux, measured in the comoving frame, generated
in the form of backward disturbances in the shell is:
\begin{equation}
  S = V_{*}W_{*} = \frac{\dot M_{b}}{4\pi R^{2}}\Gamma c^{2} \ .
  \label{eq:baf}
\end{equation}
For a fireball of initial energy $E_{0}$, this baryon loading produces only a
weak shell deceleration as long as
\begin{equation}
  t < t_{d} = \frac{3E_{0}}{\dot M_{b} \Gamma c^{2}} \ ,
  \label{eq:dec}
\end{equation}
where $t_{d}$ is the deceleration time. In other words, the shell is
still weakly perturbed as long as the mass loading rate is smaller
than $\dot M_{d} = \rho_{b} \Gamma 4\pi R_{d}^{2} c$ where $R_{d} =
ct_{d}$ is the deceleration radius. There are in fact two deceleration
radii corresponding to the two stages of baryon loading.

The relaxation of the baryon stream in the fireball is described by
the following simplified model coupling the relative level of Alfv\'en
energy density $u$ with the relative motion of the stream
$\beta_{b}-\beta_{*}$ with respect to the backward A-waves
so-generated. The coordinate $r$ denotes the radial distance from this
external sheet, increasing $r$ means approaching the shell center and
the ``initial'' condition is $\Gamma_{b}(0) = \Gamma$.
\begin{eqnarray}
    V_{*}\frac{du}{dr} & = & 2 \,g u
    \label{eq:cdw}  \\
    c\frac{d \Gamma_b}{dr} & = & - \nu \beta \Gamma_b
    \label{eq:cdb}
\end{eqnarray}
where in a quasi-linear approximation the slowing rate $\nu = \nu_0
u$, the more intense the waves the more efficient the scattering.  The
energy flux conservation implies that
\begin{equation}
    \nu_0 =
    \frac{2g \Gamma}{u_{\infty} \beta_* \beta \Gamma_b}
    \label{eq:NUG}
\end{equation}
where $u_{\infty}$ is the asymptotic value of $u$, namely $W_{*}/(B^{2}/8\pi)$
given by (\ref{eq:baf}) and thus $u_{\infty} = (B_{m}/B)^{2} <1$.

\begin{figure}[h]
  \includegraphics[width=0.8\columnwidth]{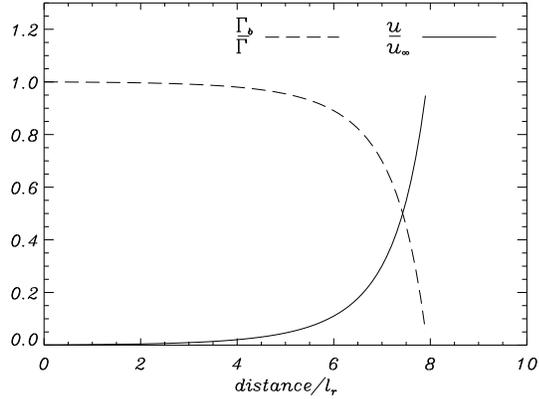}
  \caption{Stream relaxation and growth of the waves. The final
    relaxation stage, for $\beta < \beta_*$, is just a fast
    isotropisation involving plasma microphysics not detailed in this
    paper. The transition to the non-relastivistic regime gives rise
    to a stiff variation.}
 \label{fig:relax}
\end{figure}

The relaxation of the stream can be characterized by a single typical
length $l_{r} = V_{*}/g_{0}$. Fig.~\ref{fig:relax} illustrates the stream
relaxation in the shell that occurs in few $l_r$.

Let us consider the interaction with the interstellar medium.  The typical
wavelength of the excited waves, measured in light seconds ($l.sec.$), is such
that
\begin{equation}
    \lambda_{0} \simeq 10^{-3}
    \frac{\Gamma}{10^{3}}(\frac{B}{10^{3}G})^{-1} \
    l.sec. \ ,
    \label{eq:LOO}
\end{equation}
whereas the relaxation length is
\begin{equation}
    l_{r} \simeq \frac{\Gamma}{10^{3}}(\frac{n_{b}}{1 \
    cm^{-3}})^{-1/2} \
    l.sec. \ .
    \label{eq:LRE}
\end{equation}
This has to be compared with the shell width in the comoving frame;
at the deceleration time its estimate is (Rees \& M\'esz\'aros \cite{ree})
\begin{multline}
  \delta R_{d} \simeq 3 \times 10^{2} (\frac{n_{b}}{1\ 
    cm^{-3}})^{-1/3}(\frac{\Gamma}{10^{3}})^{-5/3}\nonumber\\
  \label{eq:DRD}
  \times (\frac{E}{10^{51}erg})^{1/3}(\frac{\Omega}{4\pi})^{-2/3} \ 
  l.sec..
\end{multline}
Let us consider now the interaction in the ``primary'' GRB. Expanding
$10^{-6} M_{\odot}$ within a sphere of one light-second radius
(comoving) leads to a particle density of $10^{19} cm^{-3}$. The
plasma is now collisional, but the collective process previously
developed provides a faster relaxation than both Coulomb and
pp-collision ones. The typical wavelength $\lambda_{0} \simeq 10^{-10}
\frac{\Gamma}{10^{3}}(\frac{B}{10^{10}G})^{-1} \ l.sec.$ and the
relaxation length $l_{r} \simeq 10^{-9} \frac{\Gamma}{10^{3}}
(\frac{n_{b}}{10^{18} \ cm^{-3}})^{-1/2} \ l.sec.$. Incidently, we
note that although the time is quite short, the density is so high
that the Lawson criterium for fusion is largely satisfied; moreover
the thermal energy of the particles is high ($\bar \epsilon \sim
10^{12} eV$). Thus, alpha particles and neutrons should be produced
and moreover, as mentioned previously, a significant neutrino
emission.

Once the relaxation is achieved, there are two major nonlinear
interactions with the plasma shell through magnetosonic compression.
First, intense backward A-waves that propagate almost parallel to the
magnetic field produce fast magnetosonic compression governed by the
Hada's system generalized to relativistic plasma by Pelletier \&
Marcowith (\cite{pelmar}). The transverse magnetic perturbation $b$
(reduced to the averaged field) exerts a pressure that produces fast
parallel perturbed motion $u_{\parallel}$ (specific momentum) that is
proportional to the parallel electric field (see Pelletier
\cite{pel}). This, parallel electric field is responsible for particle
acceleration or stochastic heating. For delocalized waves, particles
that resonate with the parallel electric field produce the nonlinear
Landau damping of the A-waves. The so-produced $E_{\parallel}$
efficiently injects electrons and positrons in the high energy
population.

There is also another strong nonlinear effect with the slow magnetosonic mode
(S-mode) that efficiently backscatters the primary flux of A-waves. This is
presented in the next section.

\section{Brillouin backscattering}\label{sec.bb}

As seen in the previous section, the flux of backward A-waves
becomes stronger and stronger as long as the shell is still
entraining baryons. If these waves are not be absorbed in the plasma
shell, no deceleration nor heating of the shell would stem from the
relaxation of the incoming baryons.  Moreover, Fermi acceleration with
these waves can work only if there are both forward and backward
waves. One could expect some reflection of these waves in the internal
edge of the shell; however, because they propagate at a velocity close to
the light velocity, no significant change of impedance and thus no
significant reflection occurs at the edge.  Both absorption and
backscattering efficiently develop by excitation of sound or slow
magnetosonic modes.

\subsection{Parametric Brillouin instability}

A large fraction of this flux can be backscattered (thus producing a forward
A-flux) by exciting the slow mode of the shell plasma. This is analogous to
the Brillouin backscattering, but with A-waves instead of ordinary
electromagnetic waves. The analogous Raman scattering does not work with
A-waves (three mode couplings does not work with A-waves).  A mother A-wave
($\omega_{0}, k_{0}$) of relative amplitude $b_{0}$ spontaneously gives rise
to a slow magnetosonic mode ($\omega_{s}, k_{s}$), but this S-mode is unable
to carry the whole flux of energy-momentum and a backward (forward with
respect of the shell expansion) A-wave is generated with a lower
energy-momentum ($\omega_{-}, k_{-}$) such that:
\begin{equation}
    \omega_{0} = \omega_{s} + \omega_{-} \ and \ k_{0} = k_{s} +
    k_{-} \ .
    \label{eq:bbs}
\end{equation}
For parallel propagating resonant waves,
\begin{equation}
    k_{s} = \frac{2k_{0}}{1+V_{s}/V_{*}} \ and \ k_{-} =
    k_{0}\frac{1-V_{s}/V_{*}}{1+V_{s}/V_{*}} \ ,
    \label{eq:kkk}
\end{equation}
one gets the most efficient rate of coherent wave decay:
\begin{equation}
    G_{decay} = \frac{1}{2}\sqrt{\omega_{s}\omega_{-}}
    (\frac{b_{0}^{2}}{2\beta_{p}})^{1/2} \ .
    \label{eq:decr}
\end{equation}
This result is a particular solution that can be derived from the complete
analysis made by Champeaux et al. (\cite{cham}).  For incoherent scattering,
the decay rate is lower, proportional to $b_{0}^{2}/2\beta_{p}$ instead of its
square root.

The S-modes are in turn absorbed by resonant interaction with ``thermal''
particles.  This process is clearly the most efficient to transmit the
momentum and energy of the baryon stream to the plasma shell.

Now intense forward and backward flux of long A-waves are still remaining.
There is no other way to absorb them than cosmic ray acceleration through
Fermi processes that involve resonant interaction between high energy
particles and long A-waves.

\subsection{A toy model of the backscattering process}\label{subsec.toy}

\begin{figure}[h]
  \includegraphics[width=0.8\columnwidth]{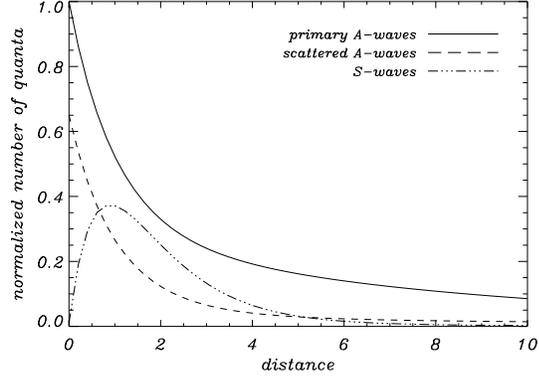}
  \caption{Solutions of the backscattering system. By order of 
    decreasing line thickness we show $N_{+}/N_{+}(0), N_{-}/N_{+}(0)$
    and $N_{s}/N_+(0)$. The distance unit is the scattering length
    defined by $V_*/w N_+(0)$. In this example $\gamma_+ = 0.1,
    \gamma_- = 0.05$ and $\gamma_s = 0.69$.}
  \label{fig:backscatter}
\end{figure}

In principle, solving the nonlinear system that governs the
backscattering process allows us to predict how the dissipated energy
is shared between heating the thermal particles of the shell and
acceleration of cosmic rays. The detailed theory involves the numbers
of quanta for each mode ($N_{+},N_{-},N_{s}$) for the mother A-waves,
the backscattered A-waves and the generated slow waves respectively),
a probability rate of scattering $w$ and absorption rate ($\gamma_{+},
\gamma_{-}, \gamma_{s}$), all these triplets depending of the three
wave vectors. In this paper we merely address a very simplified
version of the system, similar to the system obtained with random
phase approximation, but averaged over the spectral bands. We think it
would be useful to solve such a toy system to illustrate the process
and to see whether reasonable estimates can be obtained by developing
this method.  For the sake of simplicity, we assume that
backscattering occurs behind the stream relaxation sheet (in fact, it
could even start in the relaxation sheet). The system is thus the
following (in this notation a backscattered wave propagates outwards):
\begin{eqnarray}
    V_{*}\frac{\partial}{\partial r}N_{+} & = &
    -w(N_{+}N_{s}+N_{+}N_{-}-N_{-}N_{s})-\gamma_{+}N_{+}
    \label{eq:plus}  \\
    -V_{*}\frac{\partial}{\partial r}N_{-} & = &
    w(N_{+}N_{s}+N_{+}N_{-}-N_{-}N_{s})-\gamma_{-}N_{-}
    \label{eq:moins}  \\
    V_{s}\frac{\partial}{\partial r}N_{s} & = &
    w(N_{+}N_{s}+N_{+}N_{-}-N_{-}N_{s})-\gamma_{s}N_{s}
    \label{eq:son}
\end{eqnarray}
The absorption rate $\gamma_{s}$ of the S-waves by the ``thermal'' plasma is
always larger than the absorption rates $\gamma_{+}$ and $\gamma_{-}$ of the
A-waves by cosmic rays (gyro-synchrotron absorption); moreover the S-waves
have shorter wavelengths than the A-waves, the backscattered waves have even
larger wavelengthes than their mother wave and thus even less absorbed. There
are two scales in the system: a short characteristic scale associated to the
growth of the S-waves $l_{s} = V_{s}/(\gamma_{s}-G)$, where $G =
w(N_{+}(0)-N_{-}(0))$ is the nonlinear scaterring rate, and a longer scale
associated with the decay of the mother waves
$l_{*}=V_{*}(\gamma_{s}-G)/\gamma_{s}G$; $l_{s} \ll l_{*}$ when $G \ll
\gamma_{s}$. This decay length $l_{*} \sim \beta \lambda_{0}/u$.  A ratio $T$
of the incident flux is transmitted, a ratio $R$ is reflected, a ratio $A$ is
absorbed and one has:
\begin{equation}
    T+R+A = 1 \ ;
    \label{eq:TRA}
\end{equation}
and the absorption ratio is divided into a thermal one through S-waves,
$A_{s}$, and a nonthermal one through A-waves, $A_{*}$. Thus when
$V_{*}/\gamma_{+}<\delta R$, $T \simeq 0$. By solving the system, one obtains
the backscattering rate: $R= \frac{\omega_{-}N_{-}(0)}{\omega_{0}N_{+}(0)}$.
One calculates the thermal absorption ratio:
\begin{equation}
    A_{s} = \frac{\int_{0}^{\infty}
    \gamma_{s}\omega_{s}N_{s}dr}{V_{*}\omega_{0}N_{+}(0)} \ .
    \label{eq:AS}
\end{equation}
The absorption rate into cosmic rays is deduced by using (\ref{eq:TRA}).  Such
a calculation is illustrated in Fig.~\ref{fig:backscatter}. For reasonable
absorption rates, we obtain $R \simeq 0.17$, $A_{s} \simeq 0.53$ and $A_{*}
\simeq 0.30$. Of course, a more realistic calculation should be made to
predict these ratios, but this simple model indicates that, through the
backscattering process, a sizable fraction of incoming energy can be converted
into cosmic rays. In particular, a more detailed calculation would take into
account that the longer wavelength modes are less damped and thus can be
transmitted behind the shell.

The combined process of stream relaxation and Brillouin scattering
followed by energy absorption by particles occurs first during the
``primary'' stage of the relativistic expansion when the pair wind
entrains ambient baryons coming from the progenitor debris. As argued
previously, this interaction is likely to occur without shock
formation. Then after a new adiabatic and free expansion, a new stage
of interaction occurs with the interstellar medium. The ratio of the
energy dissipated by the cosmic rays acceleration over the initial
fireball energy is $E_{cr}/E = A_* R(t)^{3}/R_{d}^{3}$ before the
deceleration time; thus, at $R=0.3 R_d$, about $10\%$ of the fireball
energy in converted into cosmic rays.  When the fireball radius
reaches the deceleration radius, a strong collisionless shock is set
up.  However it can be considered as a hardening of the previous
process. Indeed, incoming protons are reflected in the relaxation
front giving rise of the external shock, whereas the backward
magnetosonic wavefronts turn to internal shocks.  The previous
description applies during the predeceleration stage.  However,
although the theory of beam-plasma instability is no longer relevant
when the shock is set up, the Alfv\'en fronts are still behind the shock
and still accelerate cosmic rays efficiently.

\section{Fermi acceleration of UHE Cosmic Rays}\label{sec.uhecr}

As stressed by Pelletier (\cite{pel}), intense forward and backward
A-waves of long wavelength (delocalized or localized) are necessary to
accelerate UHE Cosmic Rays. The intense localized fronts propagate
with a nonlinearly modified velocity such that $\gamma_{*} \mapsto
\gamma_{nl} = \gamma_{*} + \delta \gamma$ with $\delta \gamma /
\gamma_{*}\sim \beta_{*}b_{m}^{2}$ ($b_{m}$ being the maximum reduced
amplitude of the perturbed magnetic field of the front measured in its
comoving frame).  These intense wave packets scatter particles of
Larmor radius smaller than their width or wavelength in their frame in
a few gyro-periods. In a scattering time, the particles energy gains a
factor $\gamma_{*}^{2}$ and the momenta are concentrated in the front
cone of half-angle $1/\gamma_{*}$.  Further acceleration requires
other fronts propagating in the opposite direction, because wavefronts
propagating in the same direction at almost the same velocity $V_{*}$
tend to roughly isotropize the distribution of interacting particles
with respect to their comoving frame. Thus it is crucial to get both
forward and backward fronts to accelerate cosmic rays and the
Brillouin backscattering process of section \ref{sec.bb}, that
generates longwavelength perturbations in a time $\sim G^{-1}$, is the
appropriate and efficient solution. The relativistic regime of Fermi
acceleration is deeply different from the nonrelativistic regime, not
only because of the anisotropy effect but also because the
acceleration time scale can become shorter than the scattering time
scale. This is a crucial advantage that is illustrated by the
following result (Pelletier \cite{pel}) obtained in the case of a
permanent flux of both forward and backward waves, the energy gain
during a time $\Delta t$, smaller than the scattering time, is such
that
\begin{equation}
    <\Delta p^{2}> = p^{2} \beta_{*}^{2} \gamma_{*}^{2}
    \nu_{s}\Delta t \ ,
    \label{eq:dacc}
\end{equation}
where $\nu_{s}$ is the scattering frequency. Moreover, because $B>B_{eq}$,
$\gamma_{*}> 2$, and the energy jump at each scattering is large and the
statistic evolution cannot be treated by a Fokker-Planck description.

Let us give some more details about the localized fronts.  Intense
longwaves tend to self-organize in forward and backward relativistic
fronts of amplitude $b_{m} \sim (r_{*}/\xi)^{-1/2}$, where $\xi$ is
the front width and $r_{*} = (<\gamma^{2}>/\bar \gamma)
V_{*}/\omega_{c}$ is the minimum scale of relativistic MHD (Pelletier
\& Marcowith \cite{pelmar}); it is worth noting that the perturbation
amplitudes appear $\Gamma^{2}$ times larger in the observation frame.
At each crossing of these fronts, the gain is by a factor
$\gamma_{*}^{4}$ within few gyro-periods.  Such localized fronts can
be approximately described as solitons (Pelletier \& Marcowith
\cite{pelmar}; Pelletier \cite{pel}), they differ from ideal solitons
because of the Fermi acceleration that produces a kind of
Landau-synchrotron damping of them. When ideal solitons cross each
other they do not destroy, whereas damping destroys them and their
complete absorption corresponds to the strongest efficiency of the
Fermi process. A similar idea was used by Daigne \& Mochkovitch
(\cite{daim}) with internal shocks to explain the gamma emission. Here
we interpret the internal shocks as internal Alfv\'en fronts which can
contribute to short scale variations in the light curve, and possibly
more easily than pure hydrodynamic shocks.

For a bulk Lorentz factor of the shell of $10^{3}$, particles in GRBs
have to reach $10^{17} eV$ in the comoving frame in order to supply
the UHE Cosmic Rays population beyond the GZK threshold. We consider
the interaction with the interstellar medium only because, as
previously showed, synchrotron losses prevent UHE cosmic ray
production, inside a radius $R_s$ (subsection \ref{sec.mfs}). Assuming
that the largest wavelengthes are one tenth of the effective shell width
(limited by causal connection), a magnetic field just larger than the
equipartition value at deceleration radius ($B_{eq} \sim 10^{3} G$
say) is enough to get Larmor radii of that size (i.e.  $\sim 10^{-5}
pc$), which corresponds to an energy of order $10^{17} eV$. With $\bar
\gamma \sim 10^{8}$, the scale $r_{*} \sim 10 \ l.sec.$. Therefore,
few fronts having a width of $1-10 r_{*}$ that cross each other within
a shell width of $3 \times 10^{2} l.sec.$ can produce protons of
$10^{21}eV$ within one second with respect of an observer frame during
the interaction with the interstellar medium.  When the strong
relativistic shock has formed ($R>R_{d}$), external particles gain
energy by a first Fermi half-cycle by a factor $\Gamma_{s}^{2}$. They
are then injected in the relativistic A-fronts behind the shock and
suffer further acceleration up to the ultimate energy.

The local distribution is more likely quasi monoenergetic rather than
a powerlaw. However the local characteristic energy is proportional to
the product $BR$ and thus the global distribution reflects the
distribution of the product $BR$; as stated by Pelletier (\cite{pel}),
a distribution close to $\epsilon^{-2}$ is obtained for $B \propto
R^{-m}$ with $m$ close to $2$.

Our main conclusion is that Gamma-Ray Bursts are capable of producing
the UHE Cosmic Rays through a Fermi process with relativistic Alfv\'en
waves. These intense waves are generated by the two stages of baryon
entrainment, first during the primary GRB stage, second during the
interaction with the interstellar medium. The Fermi process works
because of the Brillouin backscattering process that turns out to be
appropriate and efficient, moreover it allows the heating and
deceleration of the shell plasma by the incoming flux. Although the
primary stage does not produce UHE cosmic rays because of synchrotron
losses, the Fermi acceleration allows the maintenance of a high energy proton
population between $R_h$ and $R_{pp}$ that can emit a significant low
energy neutrino flux of the order $10^{-2}E$. A high energy neutrino
flux can be generated through the p$\gamma$-process by UHE cosmic rays
at the end of the free expansion, typically a sizable fraction of the
UHE cosmic ray energy in the form of $10^{14}$--$10^{16} eV$ neutrinos
(Rachen \& M\'esz\'aros \cite{rac}).

\begin{acknowledgements}
  The authors are grateful to Y. Gallant, G. Henri, A. Marcowith, R.
  Mochkovitch and J. Rachen for fruitful discussions.
\end{acknowledgements}

\end{document}